\documentclass{moriond}

\def\be{\begin{equation}}
\def\ee{\end{equation}}
\def\bea{\begin{eqnarray}}
\def\eea{\end{eqnarray}}

\newcommand{\Photo}{
}

\begin{document}
\vspace*{4cm}
\title{STERILE NEUTRINO DARK MATTER AND LEPTOGENESIS \\
IN LEFT-RIGHT SYMMETRIC THEORIES}

\author{David Dunsky$^{1,2}$}
\author{Lawrence J. Hall$^{1,2}$}
\author{Keisuke Harigaya$^3$ (speaker)}
\address{$^1$Department of Physics, University of California, Berkeley, California 94720, USA}
\address{$^2$Theoretical Physics Group, Lawrence Berkeley National Laboratory, Berkeley, California 94720, USA}
\address{$^3$School of Natural Sciences, Institute for Advanced Study, Princeton, NJ 08540, USA}

\maketitle\abstracts{
Left-Right symmetric theories
solve the strong CP problem and explain the small Higgs quartic coupling at high energy scales via the Higgs Parity mechanism, which forces the Higgs quartic coupling to vanish at the Left-Right symmetry breaking scale. They also predict three right-handed neutrinos; one may be stable and provide dark matter, and another may decay and explain the baryon asymmetry of the universe through leptogenesis. For the dark matter abundance to arise from freeze-out, the required range of the Left-Right symmetry breaking scale is $10^{10}\mathchar`-10^{13}$ GeV, in remarkable agreement with the energy scale at which the Higgs quartic coupling vanishes. The allowed parameter space can be probed by the warmness of dark matter, precise measurements of the top quark mass and QCD coupling constant by future colliders and lattice computations, and measurement of the neutrino mass hierarchy.
}

\section{Introduction}

Left-Right gauge symmetry, $SU(2)_L \times SU(2)_R$, combined with  space-time parity, can solve the strong CP problem~\cite{Beg:1978mt,Mohapatra:1978fy}.
We consider a theory with the minimal Higgs content, where $SU(2)_{R,L}$ is broken by an $SU(2)_{R,L}$ doublet Higgs,  $H_{R,L}$. In contrast to theories with triplets and bi-fundamentals, the theory can solve the strong CP problem without introducing extra symmetries~\cite{Babu:1988mw,Babu:1989rb,Hall:2018let}. Furthermore, the Higgs Parity mechanism~\cite{Hall:2018let} can force the Standard Model (SM) Higgs quartic coupling to vanish at the Left-Right symmetry breaking scale, explaining the small Higgs quartic coupling at high energy scales.

Left-Right symmetry also predicts three right-handed neutrinos that are produced by gauge boson exchange in the early universe. One right-handed neutrino may be stable enough to be dark matter (DM), and the others may decay and produce the baryon asymmetry via leptogenesis~\cite{Fukugita:1986hr}.
In this proceeding, we discuss a freeze-out scenario where the right-handed neutrinos are in thermal equilibrium with the SM particles in the early universe and later decouple. Right-handed neutrino DM would be over-produced, but the out-of-equilibrium decay of the other two right-handed neutrinos dilutes the DM~\cite{Asaka:2006ek,Bezrukov:2009th,Nemevsek:2012cd,Dror:2020jzy}. We find that the DM abundance and successful natural leptogenesis require the Left-Right symmetry breaking scale to be in the range $10^{10}\mathchar`-10^{13}$ GeV, which agrees with the energy scale at which the Higgs quartic coupling vanishes. The remaining parameter space can be probed by the warmness of DM, precision measurements of the top quark mass and QCD coupling constant by future colliders and lattice computations, and measurement of the neutrino mass hierarchy.

\section{Left-Right symmetry and the strong CP problem}
Parity symmetry can forbid the QCD $\theta$ term. In the SM, parity symmetry is explicitly broken because the $W$ boson couples only to left-handed fermions. To impose a parity symmetry on the theory, one must introduce a new $SU(2)_R$ gauge symmetry under which the right-handed fermions are charged and a Left-Righty symmetry under which $SU(2)_L$ and $SU(2)_R$ are exchanged~\cite{Beg:1978mt,Mohapatra:1978fy}. The resultant gauge symmetry is $SU(3)_c\times SU(2)_L\times SU(2)_R\times U(1)_{B-L}$.
The correction to $\overline{\theta}$ from the phases of the quark mass matrix may be suppressed since the quark yukawa coupling is Hermitian and hence possesses real eigenvalues. There is a danger, however, of the Higgs fields obtaining complex field values generating a contribution to $\bar{\theta}$. This is in fact the case if the quark mass is given by the condensation of $SU(2)_L\times SU(2)_R$ bi-fundamental scalars, and one must impose extra symmetries to forbid the physical complex field value of the bi-fundamental scalars~\cite{Beg:1978mt,Mohapatra:1978fy,Kuchimanchi:1995rp,Mohapatra:1995xd}. 

This extra complexity is avoided if $SU(2)_R$ is broken by an $SU(2)_R$ doublet Higgs $H_R$ with $\left\langle{H_R}\right\rangle \equiv v_R$ and $SU(2)_L$ is broken by an $SU(2)_L$ doublet Higgs $H_L$ with $\left\langle{H_L}\right\rangle \equiv v$,
and the quark masses are generated by dimension-five operators
\begin{equation}
\label{eq:Lquark}
 \frac{c_{ij}^u}{M} \, q_i \bar{q}_j H_L H_R +    \frac{c_{ij}^d}{M} \, q_i \bar{q}_j H_L^\dag H_R^\dag  + {\rm h.c.},
\end{equation}
where $q_i/\bar{q_i}$ are the left/right-handed quarks. The parity symmetry forces the coefficients $c_{ij}^{u,d}$ to be Hermitian. The dimension-five operators may be generated by exchange of heavy Dirac fermions; see~\cite{Hall:2018let} for possible gauge charges of them.
It is also possible that some of the Dirac masses are small so that some of the right-handed SM fermions dominantly come from the Dirac fermions rather than the $SU(2)_R$ doublets.
Since the phases of $H_{R,L}$ can be removed by gauge transformations, no $\theta$ term is generated from the quark mass matrix. The charged lepton mass is given by similar dimension-five operators.

Note that parity does not forbid phases in the quark mass matrix, and the CKM phase is readily obtained. This is in contrast to the CP solution to the strong CP problem, where the CKM phase is forbidden by the CP symmetry and must arise from spontaneous CP breaking, reintroducing the strong CP problem unless a sophisticated mechanism is introduced~\cite{Nelson:1983zb,Barr:1984qx,Bento:1991ez}.

\section{Right-handed neutrino dark matter}
Left-Right symmetry requires right-handed partners of the three left-handed neutrinos $\nu_i$, namely, the three right-handed neutrinos, $N_i$. One of them may be stable enough to be DM. The masses of the left- and right-handed neutrinos are generated by
\begin{equation}
\label{eq:L}
\frac{c_{i}}{2M} \left( \bar{\ell}_i \bar{\ell}_i H_R H_R +  \ell_i \ell_i H_L H_L   \right)  -  \frac{b_{ij}}{M} \,  \ell_i \bar{\ell}_j  H_L H_R + {\rm h.c.},
\end{equation}
where $\ell_i/\bar{\ell_i}$ are the left/right-handed leptons. Since $v_R \gg v$, the masses of right-handed neutrinos $M_i$ are dominated by the first term. The left-handed neutrino mass is given by the second term and the combination of the first and the third term, 
\begin{equation}
\label{eq:numass}
m_{ij} \, = \, \delta_{ij} \frac{v^2}{v_R^2} M_i - y_{ik} v \; \frac{1}{M_k} \; y_{jk} v, \hspace{1in} y_{ij} \equiv b_{ij} \frac{v_R}{M}.
\end{equation}
The second term in Eq.~(\ref{eq:numass}) is the see-saw contribution~\cite{Minkowski:1977sc,Yanagida:1979as,Mohapatra:1979ia,GellMann:1980vs}.

We identify $N_1$ with DM. The cosmological production of $N_1$ is as follows:
The right-handed neutrinos are kept in thermal equilibrium with the SM bath at sufficiently high temperatures in the early universe by the $SU(2)_R\times U(1)_{B-L}$ gauge interaction and decouple at lower temperatures. The resultant $N_1$ abundance is greater than the observed DM abundance unless $M_1 < 100$ eV. Such a mass range is excluded by the Tremaine-Gunn~\cite{Tremaine:1979we,Boyarsky:2008ju,Gorbunov:2008ka} and warmness~\cite{Narayanan:2000tp,Seljak:2006qw,Irsic:2017ixq,Yeche:2017upn} bounds.
We assume that $N_2$ is long-lived, dominates the universe, and decays at a temperature $T_{\rm RH}$  to dilute $N_1$. The $N_1$ abundance is then
\begin{equation}
\frac{ \Omega_{N_1} }{ \Omega _{ {\rm DM}}}  \simeq \left( \frac{ M _1 }{ 10 \, {\rm keV} } \right) \left( \frac{ 300 \,{\rm GeV} }{ M _2 } \right) \left( \frac{ T _{\rm RH}}{ 10 \, {\rm MeV}  } \right).
\end{equation}

$N_{1}$ must be cosmologically stable and $N_{2}$ sufficiently stable to decay while dominating the energy density of the universe, putting upper bounds on $y_{1i}$ and $y_{2i}$. Moreover, $N_1$ must be light enough otherwise the required $T_{\rm RH}$ is below the BBN bound of 4 MeV~\cite{Kawasaki:1999na,Kawasaki:2000en,Hasegawa:2019jsa}, putting an upper bound on $c_{1}$. From these bounds, one can show that the SM neutrino mass matrix is dominated by the following two entries~\cite{Dror:2020jzy},
\begin{equation}
\label{eq:numass2233}
m_{22} = \frac{v^2}{v_R^2} M_2,~~m_{33} = \frac{v^2}{v_R^2} M_3 - y_{33}^2 \frac{v^2}{M_3}.
\end{equation}
One of the SM neutrinos is predicted to be much lighter than the other two. The mass of $N_2$ is then enhanced over the mass of one of the two heavier SM neutrino masses by a factor of $(v_R/v)^2$. This relation provides a strong bound on the range of $v_R$ as we will see.

\begin{figure}[tb]
    \centering
    \begin{minipage}{0.5\textwidth}
        \centering
        \includegraphics[width=.95\textwidth]{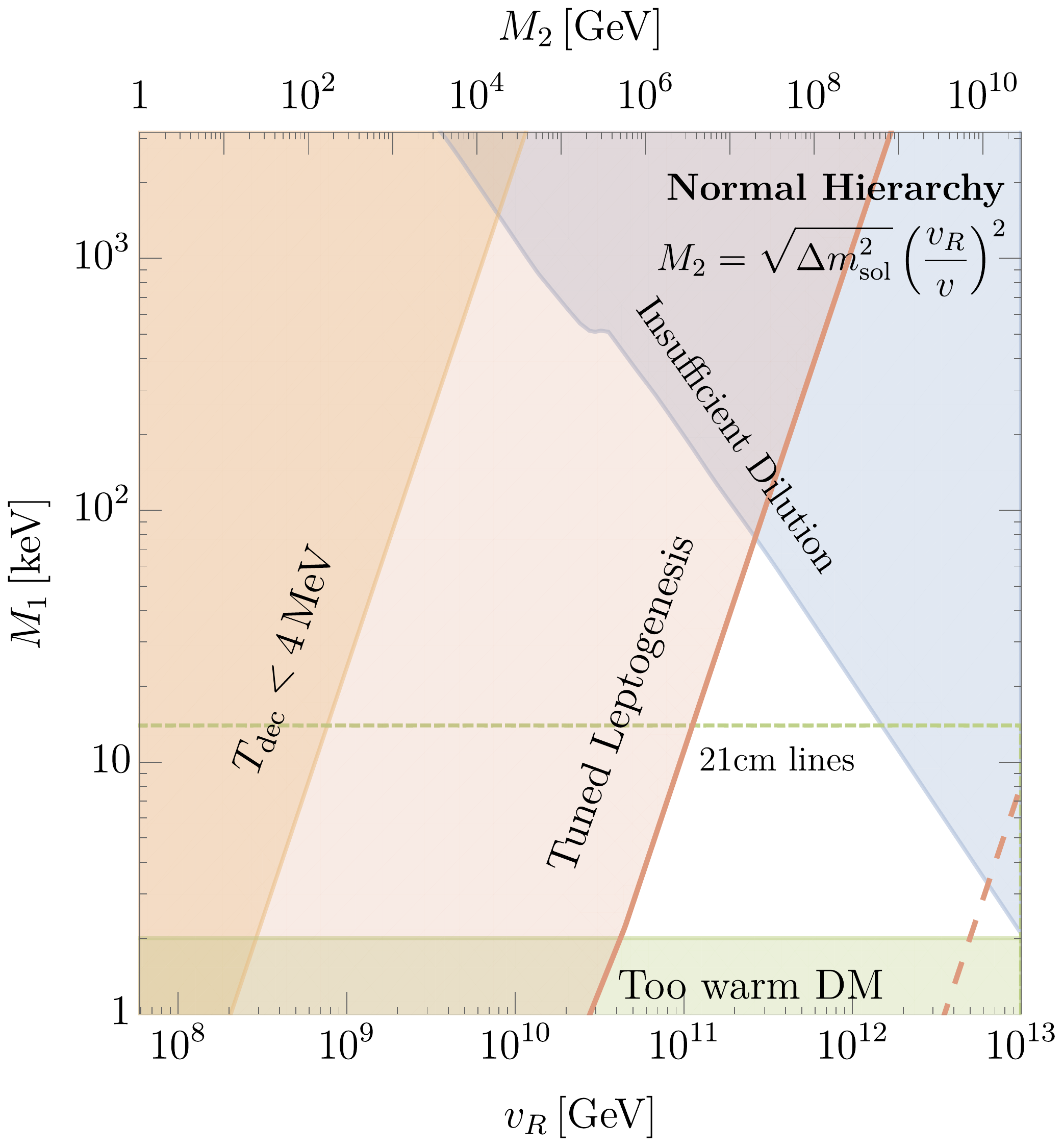} 
            \end{minipage}\hfill
    \begin{minipage}{0.5\textwidth}
        \centering
        \includegraphics[width=.95\textwidth]{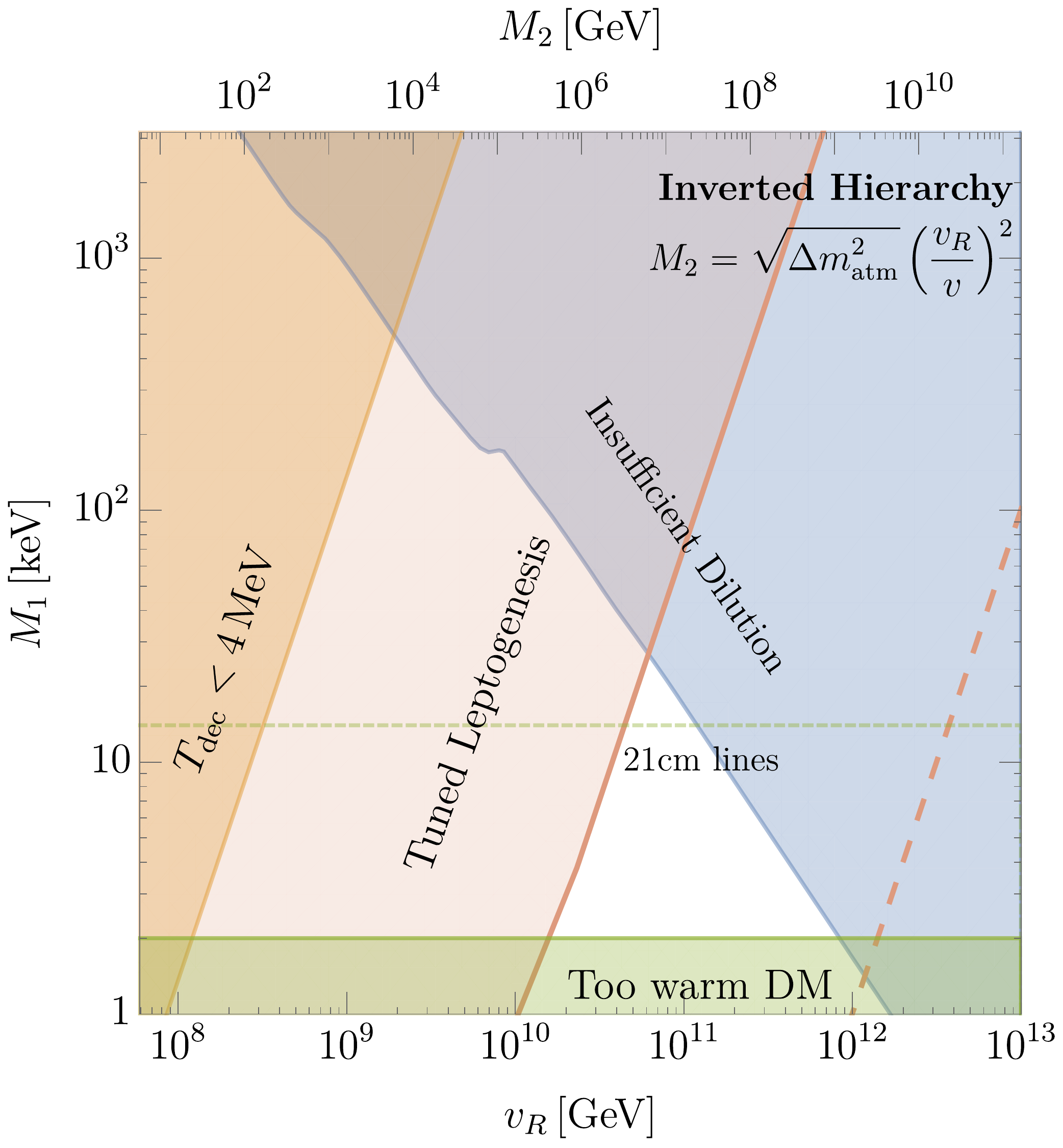} 
    \end{minipage}
    \caption{\small Constraints on the Left-Right symmetry breaking scale $v_R$ and the dark matter mass $M_1$.}
    \label{fig:vRm1}
\end{figure}

In Fig.~\ref{fig:vRm1}, we show the constraints on $v_R$ and $M_1$~\cite{Dunsky:2020dhn}. In the orange-shaded region, the required $T_{\rm RH}$ is below the BBN bound of 4 MeV. In the blue-shaded region, the decay of $N_2$ by $W_R$ exchange prevents low enough $T_{\rm RH}$. In the green-shaded region, $N_1$ is too warm.
Future observations of 21 cm lines can probe the warmness up to the green dashed line~\cite{Munoz:2019hjh}. 
We discuss the red-shaded region and the red dashed line in the next section.

\section{Leptogenesis from right-handed neutrinos}
The decay of $N_2$ into $\ell H_L$ can produce the baryon asymmetry through leptogenesis. The CP violation is provided by the interference between the tree and one-loop decay diagrams. Since $y_{1i}$ and $y_{2i}$ are required to be small, the quantum correction to the decay is dominated by $y_{33}$, and the lepton asymmetry produced per decay of $N_2$ is proportional to $y_{33}^2$.

Since $y_{33}$ contributes to the SM neutrino mass as shown in Eq.~(\ref{eq:numass}), assuming no cancellation between the two terms in $m_{33}$, $y_{33}^2$ is at most $m_{33}^2 v_R^2/ v^4$. The bound is saturated when the two terms are comparable. With this upper bound, the observed baryon asymmetry can be explained only to the right of the red dashed line in Fig.~\ref{fig:vRm1}.

The baryon asymmetry is enhanced when $M_2\simeq M_3$, leading to a resonance~\cite{Flanz:1994yx}. Such a  degeneracy can be explained by an approximate flavor symmetry. The symmetry is necessarily broken by the charged lepton mass, so the maximal natural degeneracy is $y_{\tau}^2/(8 \pi^2) \simeq 10^{-6}$. With this maximal natural enhancement, the region in Fig.~\ref{fig:vRm1} to the right of the red-shading is consistent with the observed baryon asymmetry.

The baryon asymmetry is also enhanced if a cancellation between the two terms in $m_{33}$ allows a larger $y_{33}$. Such a cancellation is natural if the dimension-five operators in Eq.~(\ref{eq:L}) result from a heavy singlet fermion $S$ with the following interactions,
\begin{equation}
 \lambda \ell_3  S H_L +  \lambda \bar{\ell}_3 S H_R + \frac{1}{2} M_{S} S^2 + {\rm h.c.}
\end{equation}
After integrating out $S$ at tree-level, only one linear combination of $\nu_3$ and $N_3$, which is dominantly $N_3$, obtains a Majorana mass and hence the SM neutrino remains massless.
This can be interpreted as a cancellation between the two terms in $m_{33}$. The allowed parameter space, however, is not as large as that achieved by $M_2\simeq M_3$~\cite{Dunsky:2020dhn}.

\section{Standard Model parameters and the Left-Right symmetry breaking scale}

An intriguing feature of the $SU(2)_R\times SU(2)_L$ breaking by $H_R$ and $H_L$ is that the symmetry breaking scale $v_R$ is predicted as a function of the SM parameters including the top quark mass and the QCD coupling~\cite{Hall:2018let,Dunsky:2020dhn}. The scalar potential of $H_R$ and $H_L$ is
\begin{equation}
V(H_R,H_L) = - m  ^2 \left( \left| H_R \right|^2  + \left| H_L \right|^2 \right) + \frac{\lambda}{2} \left( \left| H_R \right|^2  + \left| H_L \right|^2 \right) ^2 + \lambda' \left| H_R \right|^2 \left| H_L \right|^2.
\end{equation}
We assume that $m^2$ is positive and much larger than $v$. $H_R$ obtains a large vacuum expectation value, $m/\lambda^{1/2}$, and spontaneously breaks the Left-Right symmetry. After integrating out $H_R$ at tree-level around this vacuum, the low energy effective potential of $H_L$ is
\begin{equation}
V_{\rm LE}(H) = \lambda' \; v'^2  \;  \left| H_L \right|^2 - \lambda' \left(1  + \frac{\lambda'}{2 \lambda} \right) \left| H_L \right|^4.
\end{equation}
The hierarchy $v\ll v_R$ is obtained only if the quadratic term is small, which requires $|\lambda'| \ll 1$. Then the quartic coupling of $H_L$ is also enforced to be very small at the energy scale $v_R$. Quantum corrections, dominated by renormalization group running in the SM effective field theory between $v_R$ and $v$, generate a positive SM Higgs quartic coupling at the weak scale $\lambda_{\rm SM}(v)\simeq 0.1$. From the perspective of running from low to high energy scales, the energy scale at which the SM Higgs quartic coupling nearly vanishes is the scale $v_R$. We call this the ``Higgs Parity mechanism". Threshold corrections to $\lambda_{\rm SM}(v_R)$ are given in~\cite{Dunsky:2020dhn,Hall:2019qwx}.

\begin{figure}[tb]
    \centering
    \begin{minipage}{0.5\textwidth}
        \centering
        \includegraphics[width=.95\textwidth]{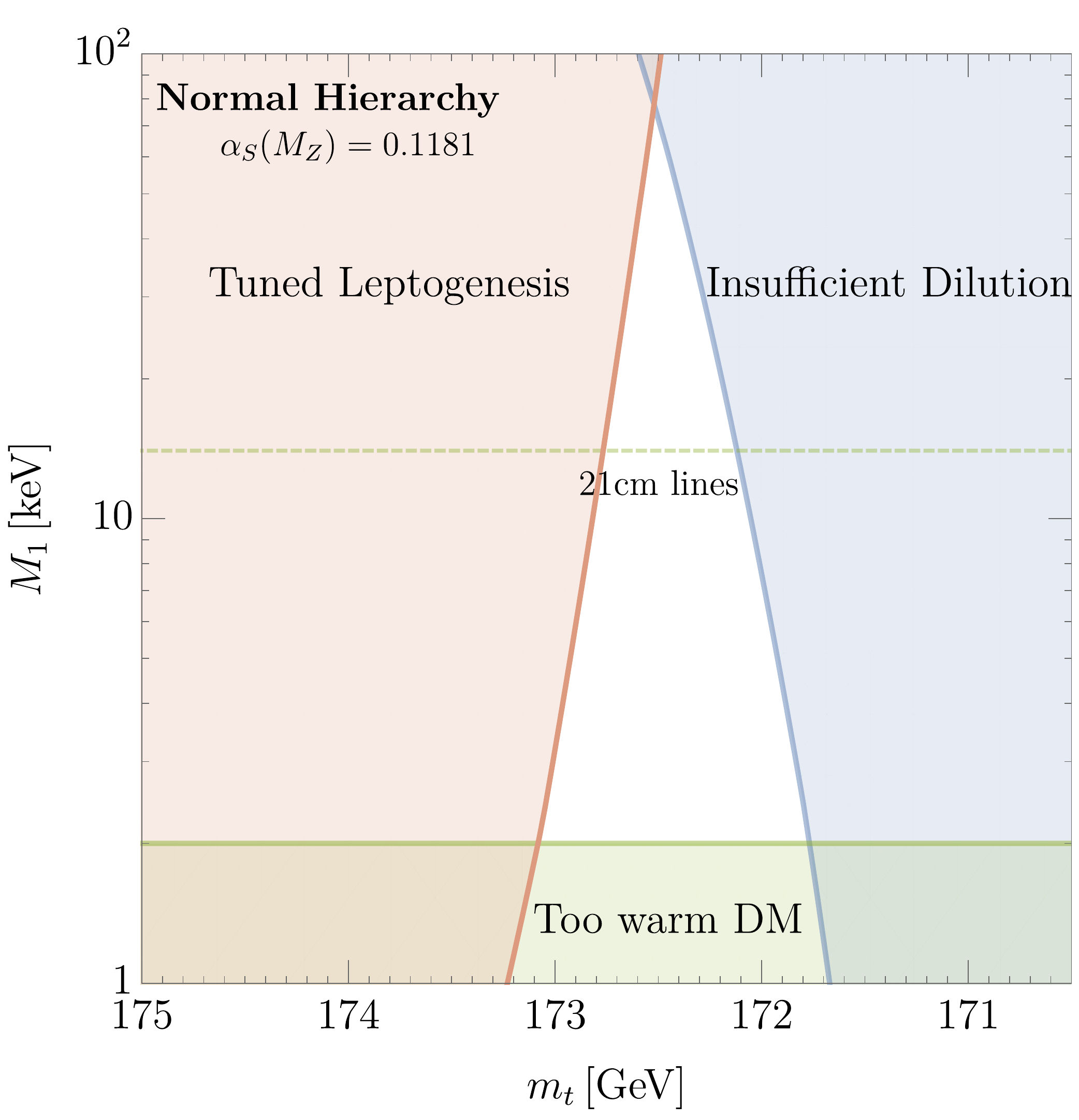} 
            \end{minipage}\hfill
    \begin{minipage}{0.5\textwidth}
        \centering
        \includegraphics[width=.95\textwidth]{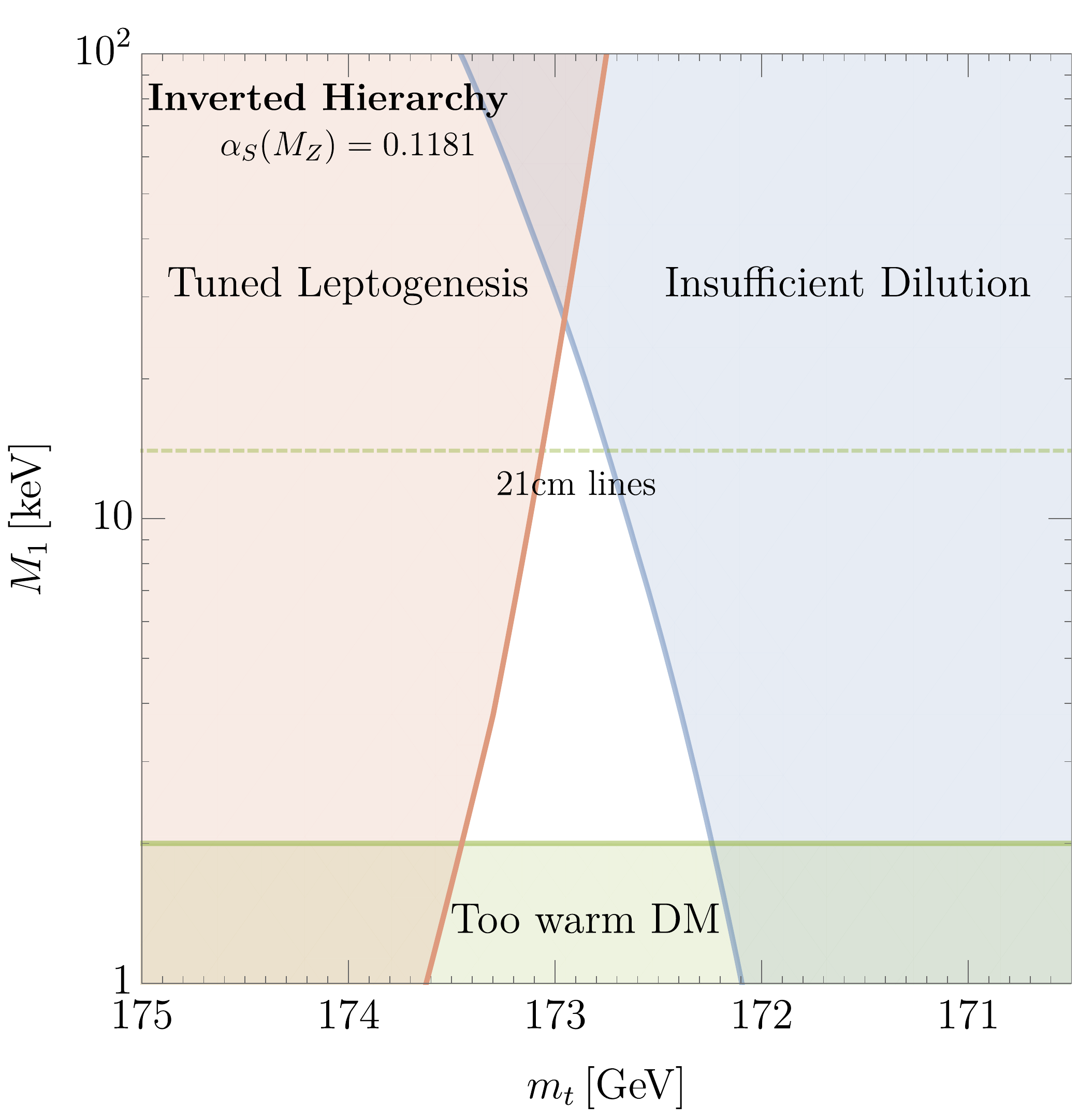} 
    \end{minipage}
    \caption{\small Constraints on the top quark mass $m_t$ and the dark matter mass $M_1$.}
    \label{fig:mtm1}
\end{figure}

Since the running of $\lambda_{\rm SM}$ is determined by SM parameters, especially the top quark yukawa and QCD coupling constants, $v_R$ is predicted as a function of these, and vice versa.
In Fig.~\ref{fig:mtm1}, we recast the constraints of Fig.~\ref{fig:vRm1} into those on the top quark mass $m_t$ and the DM mass $M_1$, for a fixed QCD coupling constant. Remarkably, the allowed range of $v_R$ for $N_1$ DM and leptogenesis from $N_2$ is consistent with the observed top quark mass, $m_t = (172.76 \pm 0.30)$ GeV.
The parameter space can be further probed by the warmness of DM, precision measurements of the top quark mass and QCD coupling constant by future colliders~\cite{Seidel:2013sqa,Horiguchi:2013wra,Kiyo:2015ooa,Beneke:2015kwa,Gomez-Ceballos:2013zzn} and lattice computations~\cite{Lepage:2014fla}, and measurement of the neutrino mass hierarchy.

\section{Discussion}

The lightness and stability of $N_1$ may be disturbed by quantum corrections from the charged fermion yukawa couplings. These corrections are sufficiently suppressed under certain conditions on the UV completion of the dimension-five operators in Eqs.~(\ref{eq:Lquark}) and (\ref{eq:L}); see~\cite{Dunsky:2020dhn} for details.

For a low enough reheating temperature after inflation, the $N_1$ abundance is set by freeze-in rather than by freeze-out and dilution by $N_2$,~\cite{Dror:2020jzy,Dunsky:2020dhn,Khalil:2008kp}. See~\cite{Dror:2020jzy,Dunsky:2020dhn} for the analysis of this scenario in the context of Left-Right symmetry.

The Higgs parity mechanism is applicable to a variety of theories where the SM Higgs has a $Z_2$ partner. Such theories provide correlations between SM parameters and the proton decay rate~\cite{Hall:2019qwx}, the DM direct detection rate~\cite{Dunsky:2019api}, and gravitational waves and dark radiation~\cite{Dunsky:2019upk}.

\section*{Acknowledgments}
This work was supported in part by the Director, Office of Science, Office of High Energy and Nuclear Physics, of the US Department of Energy under Contracts DE-AC02-05CH11231 (LJH), by the National Science Foundation under grant PHY-1915314 (LJH), as well as by Friends of the Institute for Advanced Study (KH).

\end{document}